\documentclass[12pt,preprint,letterpaper]{aastex}
\usepackage{url}

\newcommand{\sat}[1]{\it\uppercase{#1}\rm}
\newcommand{\fig}[1]{Figure~\ref{#1}}

\newcommand{\aspeed}[1]{$\sim\,$#1 km~s${}^{-1}$}

\begin{document}

\shorttitle{Current Sheet} %

\shortauthors{Liu et al.}

\title{Sigmoid-to-Flux-Rope Transition Leading to A Loop-Like Coronal Mass Ejection}

\author{Rui Liu\altaffilmark{1}, Chang Liu\altaffilmark{1}, Shuo Wang\altaffilmark{1}, Na Deng\altaffilmark{1,2}, and Haimin Wang\altaffilmark{1}}

\altaffiltext{1}{Space Weather Research Laboratory, Center for Solar-Terrestrial Research, NJIT,
Newark, NJ 07102, USA; rui.liu@njit.edu}

\altaffiltext{2}{Physics and Astronomy Department, California State University Northridge,
Northridge, CA 91330, USA}

\begin{abstract}

Sigmoids are one of the most important precursor structures for solar eruptions. In this Letter, we
study a sigmoid eruption on 2010 August 1 with EUV data obtained by the Atmospheric Imaging
Assembly (AIA) on board the Solar Dynamic Observatory (\sat{sdo}). In AIA 94~\AA\ (\ion{Fe}{18}; 6
MK), topological reconfiguration due to tether-cutting reconnection is unambiguously observed for
the first time, i.e., two opposite J-shaped loops reconnect to form a continuous S-shaped loop,
whose central portion is dipped and aligned along the magnetic polarity inversion line (PIL), and a
compact loop crossing the PIL. A causal relationship between photospheric flows and coronal
tether-cutting reconnections is evidenced by the detection of persistent converging flows toward
the PIL using line-of-sight magnetograms obtained by the Helioseismic and Magnetic Imager (HMI) on
board \sat{sdo}. The S-shaped loop remains in quasi-equilibrium in the lower corona for about 50
minutes, with the central dipped portion rising slowly at \aspeed{10}. The speed then increases to
\aspeed{60} about 10 minutes prior to the onset of a \sat{goes}-class C3.2 flare, as the S-shaped
loop speeds up its transformation into an arch-shaped loop, which eventually leads to a loop-like
coronal mass ejection (CME). The AIA observations combined with H$\alpha$ filtergrams as well as
hard X-ray (HXR) imaging and spectroscopy are consistent with most flare loops being formed by
reconnection of the stretched legs of less-sheared J-shaped loops that envelopes the rising flux
rope, in agreement with the standard tether-cutting scenario.


\end{abstract}


\section{Introduction}
CMEs and flares are responsible for the most energetic space weather phenomena, yet our knowledge
of the driving mechanisms and the actual conditions leading to the eruption is limited. To achieve
better understanding, we must identify the key aspects of eruptive structures that can provide
diagnostic information for comparison with and incorporation into the models.

A distinctive set of structures that have been under special scrutiny are sigmoids \citep{rk96},
which are forward or inverse S-shaped coronal loops seen often in soft X-ray (SXR) and sometimes in
EUV \citep[e.g.,][]{liu07} emission. They are often composed of two opposite J-like bundles of
loops which collectively make an S-shape appearance \citep{canfield07, mc08}. Sigmoidal regions are
significantly more likely to be eruptive than non-sigmoidal regions \citep{hudson98, chm99,
glover00}. The eruption is usually followed by the formation of unsheared arcades or cusped loops,
a process termed ``sigmoid-to-arcade'' evolution \citep[e.g.][]{sterling00, moore01, pevtsov02,
gibson02}. In close spatial association with sigmoids are filaments \citep{pevtsov02}, with which
the central portion of the sigmoid is approximately aligned. The filament channel is known as a
tracer of sheared field \citep{martin98}, and the dense filament material is suggested to be
supported either by a helically coiled field \citep[][and references therein]{gf06}, or by a
sheared arcade \citep[e.g.,][]{ak91}, against the solar gravity.

It is appealing to associate sigmoids with kinked flux ropes \citep{rk94, rk96}, whereas it remains
controversial whether the observed sigmoids carry sufficient magnetic twist for the onset of the
kink instability \citep{leamon03, lfb05}. \citet{td99} suggested that the transient sigmoidal
brightening outlines magnetic structures with enhanced current density. This has been demonstrated
in several MHD simulations, in which, for example, the sigmoid is identified either with separatrix
surface associated with a bald-patch topology \citep{fg04}, or with J-shaped field lines that pass
through the vertical current sheet below a rising kink-unstable loop \citep{ktt04}.

Based on morphological changes of flaring structures, a physical picture for sigmoid eruption was
proposed by \citet{ml80} and further elaborated by \citet{moore01}, in which the magnetic explosion
is unleashed by the so-called tether-cutting reconnection, occurring between the inner legs of the
two elbows of the sigmoid where they shear past each other under the filament. This reconnection
produces a shorter low-lying loop across the PIL and a longer twisted loop connecting the far ends
of the sigmoid. The eruption begins when the twisted loop becomes unstable and distends the
envelope field that overarches the sigmoid. The legs of the stretched envelope field subsequently
reconnect back to form an arcade structure and the ejecting plasmoid escapes as a CME. Evidence for
the tether-cutting model can also be found in recent observational studies \citep[e.g.,][]{wang06,
yurchyshyn06, liu07}. \citet{liu08} studied an EUV sigmoid, which is discontinuous at where it
crosses the filament prior to the eruption, but becomes a single continuous structure lying above
the filament at the eruption onset. This particular feature is at variance with \citet{moore01}, in
which the tether-cutting reconnection occurs below the filament.

The aforementioned observational studies, however, are often compromised by instrumental
constraints, due to which some important aspects of the tether-cutting model remain ambiguous: (1)
Is the eruption initiated by tether cutting which leads to the rising of a flux rope or by the loss
of equilibrium of the flux rope which later induces tether cutting below it? (2) Is the flux rope
pre-existent or formed by tether cutting? (3) Is the filament supported by the sheared legs of the
sigmoid or by an embedded flux rope? (4) Does the sigmoid itself, or the overlying arcade, or a
pre-existent flux rope, erupt as the CME?

With the launch of \sat{sdo}, we are now in a better position to address the above issues. Here we
investigate a sigmoid eruption observed by \sat{sdo}, in which a twisted flux rope formed via
tether cutting, and subsequently transformed into an arch-shaped loop distending the overlying
field and resulting in a \sat{goes}-class C3.2 flare and a loop-like CME. In the rest of the
Letter, observations and data analysis are presented in Section 2; important conclusions and
remaining issues are summarized in Section 3.

\section{Observations \& Data Analysis}

\subsection{Instruments \& Data Sets}
\begin{figure}\epsscale{0.7}
\plotone{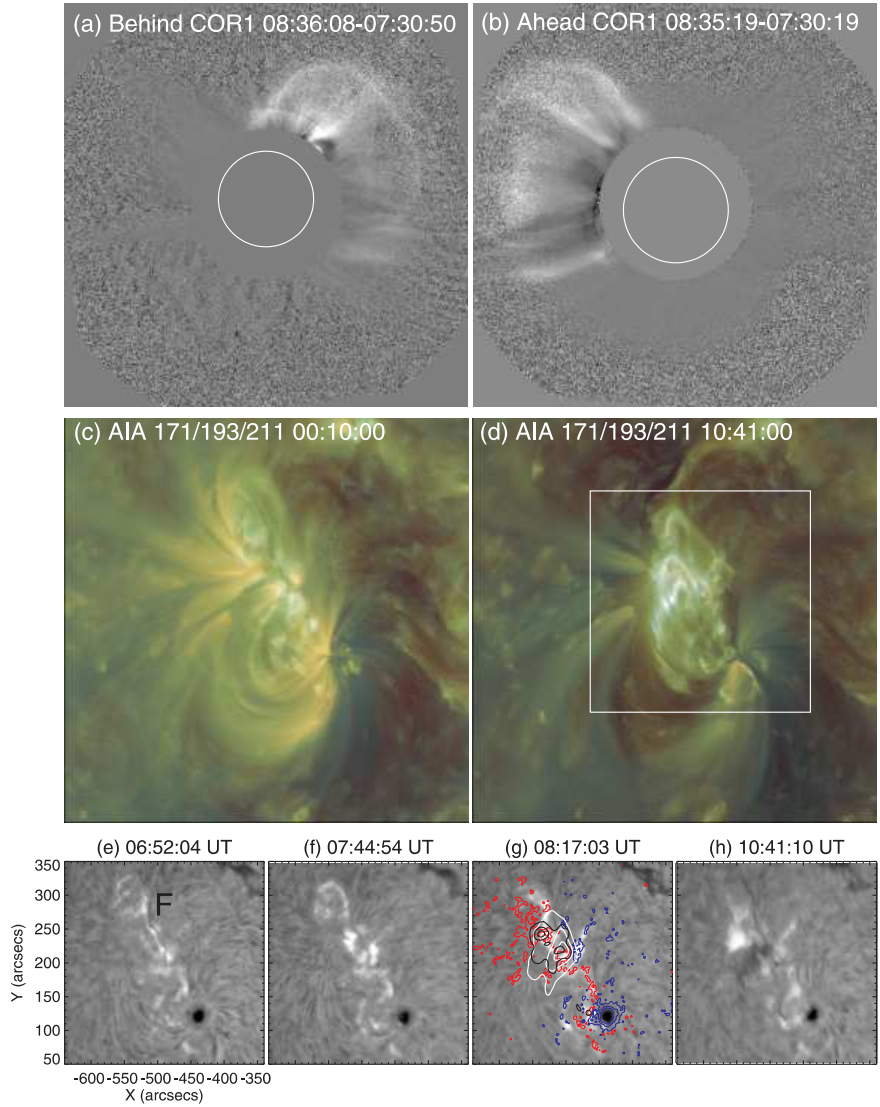} \caption{Overview of the sigmoid eruption on 2010 August 1. Panels (a) and (b)
show the CME captured by the COR1 coronagraph aboard \sat{stereo}. Panels (c) and (d) show a pre-
and a post-eruption AIA composite image, respectively. The rectangle in Panel (d) indicates the
field of view of KSO H$\alpha$ images in Panels (e)--(h). The H$\alpha$ images are registered with
the one at 08:17 UT (Panel (g)), which is co-aligned with and overlaid by the HMI magnetogram taken
at the same time. The contours levels are $\pm$100, $\pm$500 and $\pm$1000 G, with red (blue) color
for positive (negative) polarity. A filament located along the PIL is labeled ``F'' in Panel (e).
Panel (g) is also overlaid by \sat{rhessi} HXR sources integrated from 08:17 to 08:18 UT, at 3--9
keV (white) and at 12--25 keV (black). The contour levels are 50\%, 70\% and 90\% of the maximum
brightness. \label{cme}}
\end{figure}

The sigmoid studied here erupted at about 08:00 UT on 2010 August 1, and was fully covered by AIA
and HMI aboard \sat{sdo} (\fig{cme}(c) and (d)), as well as by the H$\alpha$ telescope at the
Kanzelh\"{o}he Solar Observatory (KSO; \fig{cme}(e)--(h)) starting from 06:52 UT. The subsequent
CME was detected by the COR1 coronagraph aboard both the ``Ahead'' and ``Behind'' satellites of the
Solar Terrestrial Relations Observatory (\sat{stereo}; \fig{cme}(a) and (b)), which were separated
by about 150 deg. The flare associated with the CME was detected by the Reuven Ramaty High-Energy
Solar Spectroscopic Imager \citep[\it RHESSI\rm;][]{lin02}.


AIA takes multi-wavelength images at $1''.2$ resolution and 12-second cadence. To study flaring
plasma, we focus on the 94 {\AA} filter (\ion{Fe}{18}; $\log T=6.8$), as well as the 131 {\AA}
filter which covers \ion{Fe}{8} ($\log T=5.6$), \ion{Fe}{20} ($\log T=7.1$) and \ion{Fe}{23} ($\log
T=7.2$). To enhance signal-to-noise ratio, we add every five 94~\AA\ images to yield new images at
effectively 1-min cadence. Other filters in 171~\AA\ (\ion{Fe}{9}; $\log T=5.8$), 193~\AA\
(\ion{Fe}{12}; $\log T=6.1$), 211~\AA\ (\ion{Fe}{14}; $\log T=6.3$), and 335~\AA\ (\ion{Fe}{16};
$\log T=6.4$) provide a multi-thermal perspective. The 193~\AA\ band is also sensitive to 20 MK
plasma (\ion{Fe}{24}). HMI observes the Sun at 6173~\AA\ at $1''$ resolution, and has begun to
publish line-of-sight magnetograms at 45-second cadence with a precision of 10 G.


\subsection{The Formation of the Flux Rope}
\begin{figure}\epsscale{0.9}
\plotone{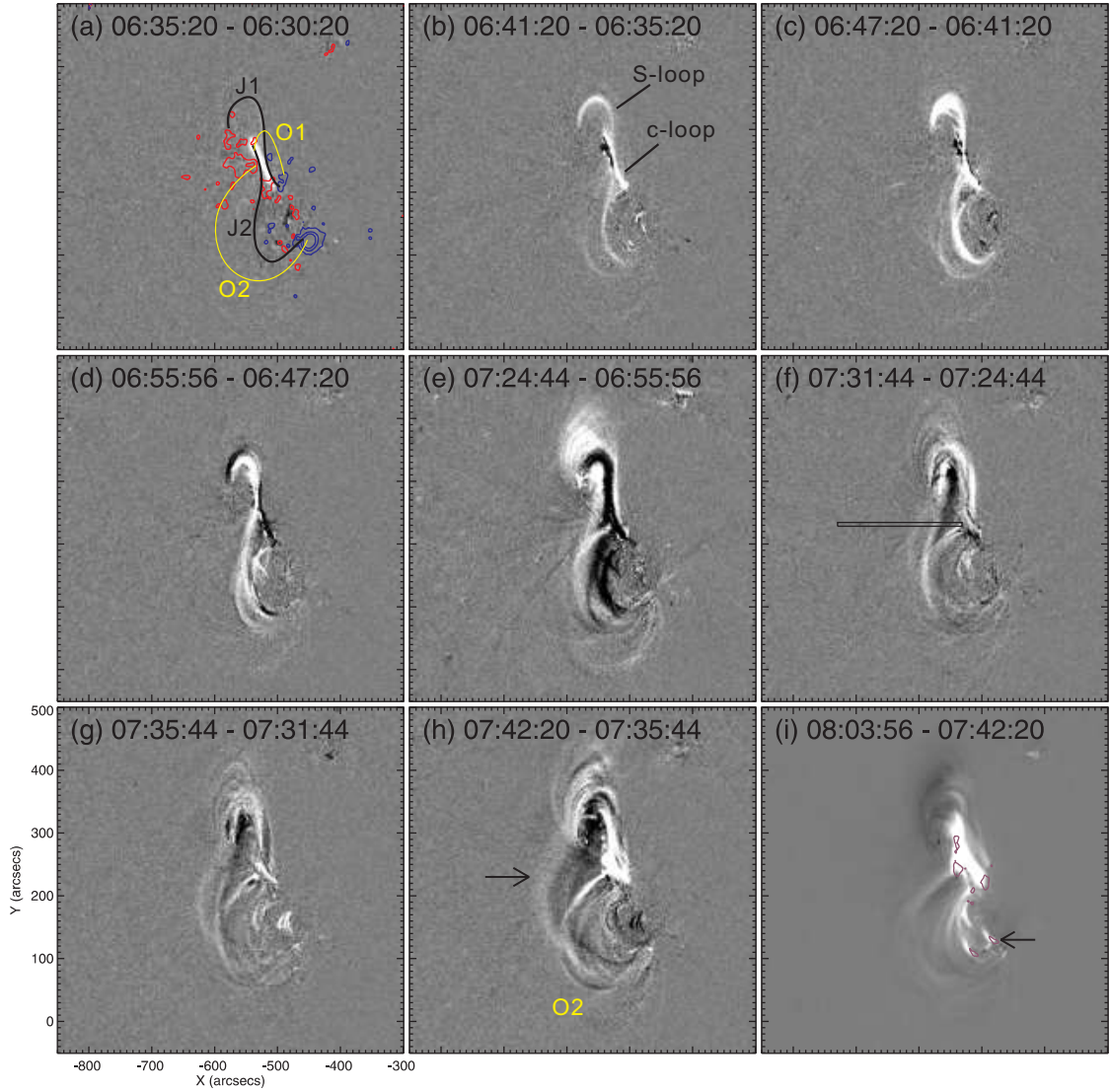} \caption{Running difference of AIA 94~\AA\ images. All images are registered
with the image taken at about 06:00 UT. Panel (a) is overlaid by an HMI magnetogram at about 06:00
UT, with the same color codes and contour levels as in \fig{cme}(g). Panel (i) is overlaid with
contours of H$\alpha$ emission (purple) indicating flare ribbons. An animation of AIA 94 \AA\ and
corresponding base-difference images is available in the online Journal. \label{diff}}
\end{figure}


The sigmoid eruption conforms to the classical ``sigmoid-to-arcade'' transformation. \fig{cme}(c)
show the sigmoid structure observed 8 hours before the eruption, which consisted of two bundles of
J-shaped loops. The arcade structure after the eruption is shown in \fig{cme}(d). The detailed
transformation is presented in \fig{diff} as the running difference of AIA 94~\AA\ images. At
06:35~UT (\fig{diff}(a)), a brightening structure appeared in the center of the active region,
which seems to be a low-lying, compact loop (hereafter c-loop) across the PIL. About five minutes
later, an inverse S-shaped loop (hereafter S-loop) started to glow, whose central portion was
apparently dipped, aligned along the PIL, and crossed the c-loop (\fig{diff}(b)). The S-loop axis
is apparently twisted by about half a turn (referred to also as flux rope hereafter), but the
individual field line could be twisted more than that.

\begin{figure}\epsscale{0.9}
\plotone{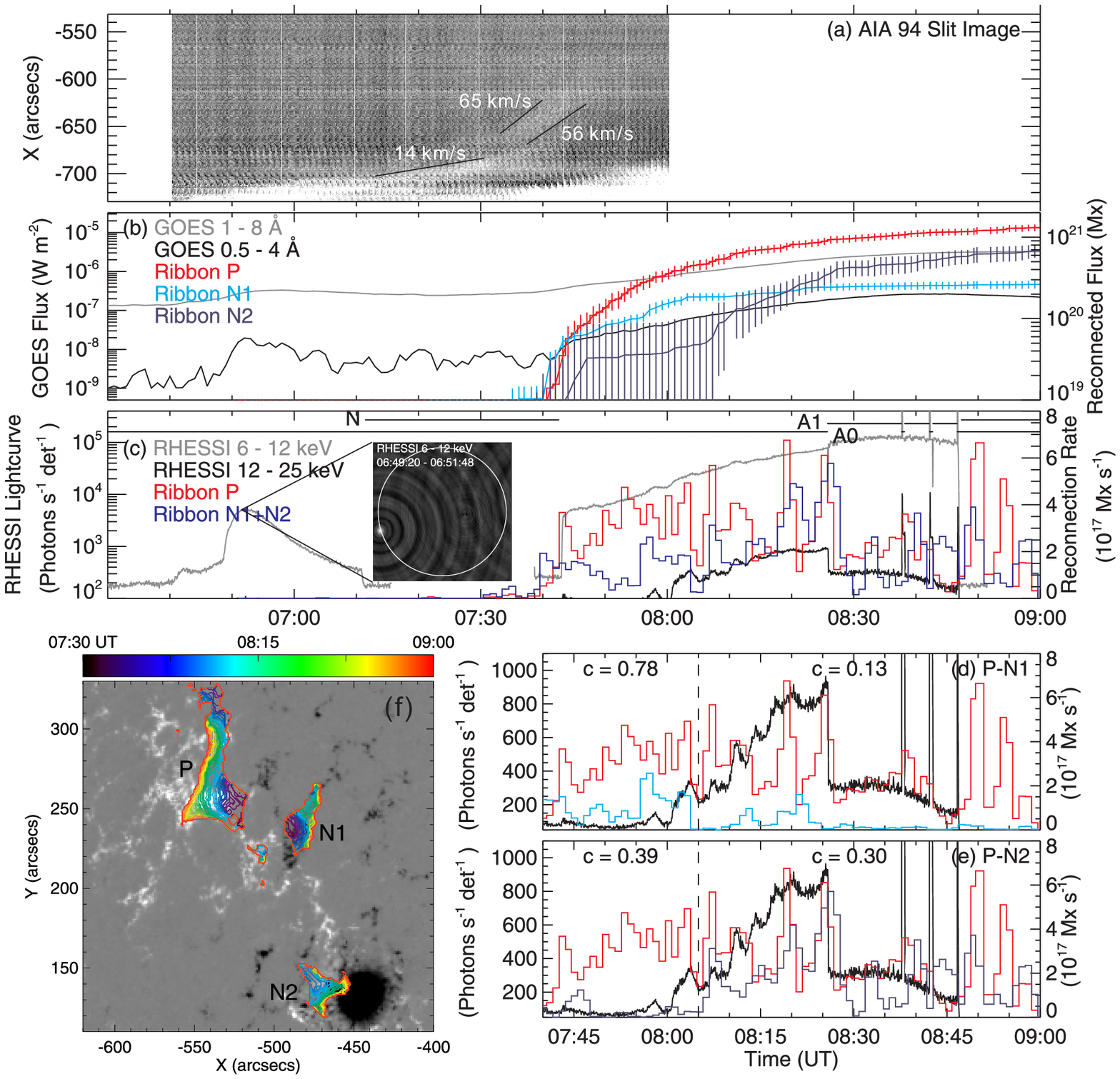} \caption{Time evolution of the sigmoid eruption in various wavelengths. In
Panel (c) the bars at the top indicate \sat{RHESSI} eclipse (`N'), and the attenuator states, `A1'
and `A0', respectively. The inset is a \sat{rhessi} Back-Projection image at 6--12 keV showing the
B3.3 flare location. The correlation coefficients between Ribbons P and N1 before and after 08:05
UT (indicated by dashed line) are shown at the top of Panel (d), and those between Ribbons P and N2
are shown in Panel (e). In Panel (f), color-coded contours display the time evolution of H$\alpha$
ribbons, overlaid on an HMI magnetogram. \label{lc}}
\end{figure}

The c-loop was soon darkened in the running difference image (\fig{diff}(d)) and became visible by
06:55 UT in cooler AIA filters, such as 335~\AA\ and 211{\AA}. The S-loop can only be clearly seen
in AIA 94~{\AA} throughout its lifetime. Its temperature is therefore well constrained, i.e., about
6~MK. Since both the S-loop and c-loop were newly formed, and of high temperature, we suggest that
they were the product of the reconnection between two J-shaped loops (hereafter J-loops) as
illustrated in \fig{diff}(a) (labeled ``J1'' and ``J2''), same as the tether-cutting reconnection
envisaged by \citet{moore01}. However, the cooling of the c-loop and the concurrent detachment of
the S-loop with it suggests the halt of the reconnection by then. The fact that the c-loop
brightened before the S-loop formed excludes the possibility that the reconnection was induced by
the flux rope. Despite the heating observed in AIA, the amount of the energy released by tether
cutting is below the detection threshold of both \sat{goes} and \sat{rhessi}, since the only flare
detected at that time was above the southeast limb (\sat{goes}-class B3.3; see the inset of
\fig{lc}(c)). This concurs with MHD simulations \citep[e.g.,][]{aulanier10}.

\begin{figure}\epsscale{0.9}
\plotone{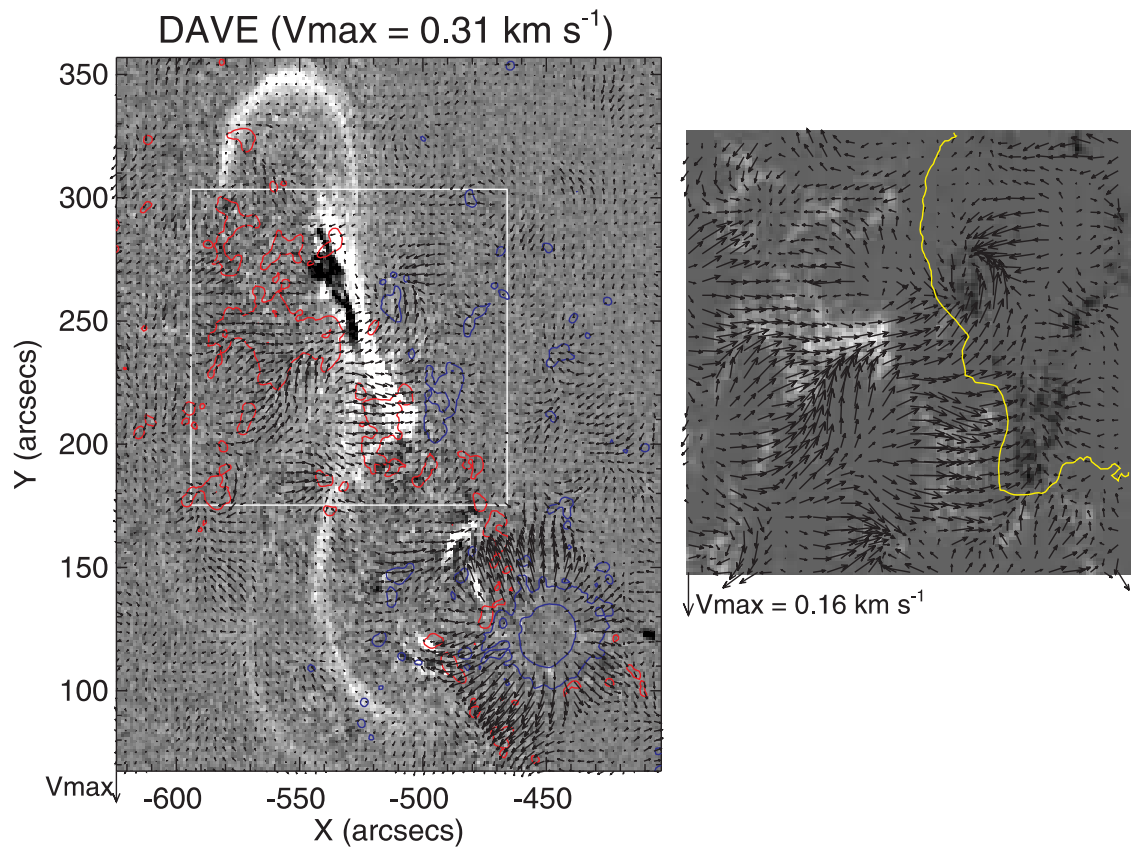} \caption{Flow field calculated with the DAVE method. \fig{diff}(b) is adopted as
the background of the left panel. Contours levels are $\pm$100 and $\pm$500 G, given by the HMI
magnetogram at 06:00 UT (background of the right panel), to which all other magnetograms used for
the calculation are registered. The rectangle in the left panel indicates the field of view of the
right panel, in which the yellow contour indicates the PIL. An animation of the HMI magnetograms is
available in the online Journal. \label{flow}}
\end{figure}

To investigate the role of photospheric flows in the formation of the flux rope, we use a series of
HMI magnetograms at 12-min cadence from 00:00 UT to 12:00 UT. The average flow field calculated by
the differential affine velocity estimator \citep[DAVE;][]{schuck05} is shown in \fig{flow}. The
most visible flow pattern is the radial outflows in the sunspot penumbra (left panel), known as
moat flows. Apart from that, one can see persistent converging motions toward, and less significant
shear motions along, the PIL (right panel). Magnetic elements of positive polarity (white patches
and red contours) show stronger converging motions than those of negative polarity (black patches
and blue contours). The overall flow pattern is independently confirmed by local correlation
tracking \citep[LCT;][]{ns88}. Flux cancellation can be seen in the animation of magnetograms, when
the white patches move into black ones. From the left panel of \fig{flow}, one can also see that
the c-loop is clearly associated with the converging flows, which supports the conjecture of
\citet{moore01}. Formation and eruption of flux ropes driven by converging flows have also been
found in 3D MHD simulations \citep[e.g.,][]{amari03,aulanier10}.

\subsection{The Eruption}
In \fig{lc}(a) we present the chronological observation of the S-loop through a slit across its
dipped portion (\fig{diff}(f)). This slit ``image'' is composed of strips cut from the AIA 94~{\AA}
base-difference images (subtracted by the image at 06:00 UT). The projected speed of the rising
flux rope can then be estimated from the brightening patterns. \fig{diff} and \fig{lc}(a) together
show that after its formation, the S-loop began to rise slowly at \aspeed{10}, driven presumably by
the upward-pointing magnetic tension force at the dip. The rising speed then increased to
\aspeed{60} at about 07:30 UT (\fig{lc}(a)), ten minutes prior to the C3.2 flare (\fig{lc}(b)). At
the flare onset (about 07:42 UT; \fig{lc}(b)), the twisted flux rope had transformed into an
arch-shaped loop (\fig{diff}(h), marked by an arrow). Snapshots of the rising flux rope may have
been captured previously \citep{moore01,mc08}. This rising and inflating loop became too diffused
to be seen after 07:50 UT, but only minutes later a CME with a similar loop morphology and loop
orientation (\fig{cme}(a) and (b)) was observed first by the \sat{stereo} ``Ahead'' Satellite from
07:55 UT onward, and then by the ``Behind'' satellite from 08:11 UT onward. No high-density
filament material (bright core) can be seen inside the CME and the filament lying along the PIL
(see \fig{cme}(e)--(h); labeled `F' in \fig{cme}(e)) remained largely intact throughout the
eruptive process. Hence the tether-cutting reconnection must have taken place above the filament
\citep[see also][]{pevtsov02, liu07, liu08}.

With the onset of the C3.2 flare, flaring loops which were less sheared than the c-loop became
visible in the central region (\fig{diff}(h) and (i)). We suggest that the flaring is due to
reconnection driven by the rising flux rope which stretched the loops overlying the central region
and resulted in the formation of a current sheet underneath, similar to the schematic diagram in
\citet[][their Figure 1]{moore01}. In \fig{diff}(a), such an overlying loop is illustrated in
yellow (labeled ``O1''). Note, however, that O1 was not visible in any AIA filter prior to the
flare. In fact, coronal loops arched over the rising flux rope were similar in magnetic
connectivity to J1 and J2 in \fig{diff}(a), except that they were located higher and less sheared.
Such an overlying loop is illustrated in yellow in \fig{diff}(a) (labeled ``O2''). One can see that
a group of loops similar to O2 were pushed upward by the rising flux rope, exhibiting a bright
outer rim and a dark inner rim in the running difference image (\fig{diff}(h)). At 08:03 UT, a
brightening ribbon in H$\alpha$ became visible near the sunspot, co-spatial with a brightening
front moving westward in AIA 94~{\AA} (\fig{diff}(i), marked by an arrow). We suggest that from
that time onward, the reconnection was due to the stretch of O2.

This interpretation is supported by H$\alpha$ observations, in which three flare ribbons are
identified and referred to as Ribbons P, N1, and N2 hereafter (\fig{lc}(f)), with P for positive
and N for negative polarity. In \fig{lc}(f) color-coded contours display the time evolution of
H$\alpha$ ribbons, overlaid on an HMI magnetogram to which H$\alpha$ images are co-aligned.
Obviously Ribbons P and N1 are connected by loops like O1, while P and N2 by loops like O2. We
calculated the reconnection rate and the reconnected flux following \citet{qy05}. The reconnection
rate is defined as the change rate of the magnetic flux passing through the newly brightened flare
area and its time integration gives the reconnected flux. The results are shown in
\fig{lc}(b)--(e). One can see that the reconnected flux at Ribbon N1 became flattened after about
08:05 UT, when the reconnection flux at Ribbon N2 started to pick up. Accordingly, the reconnection
rate at Ribbon N2 dominated over that at Ribbon N1 after 08:05 UT (\fig{lc}(b)--(e)). Moreover,
before 08:05 UT, Ribbon P was better correlated with Ribbon N1 in terms of the reconnection rate,
while after 08:05 UT, it was better correlated with Ribbon N2 (\fig{lc}(d)--(e)). About two hours
later, post-flare loops began to appear in H$\alpha$ as absorption features (\fig{cme}(h)). Like in
EUV (\fig{cme}(d)), the H$\alpha$ post-flare loops were oriented in a northeast-southwest
direction, rather than the east-west direction crossing the filament, which also indicates that the
majority of the flare loops connected Ribbons P to N2, rather than Ribbons P to N1.

\begin{figure}\epsscale{0.9}
\plotone{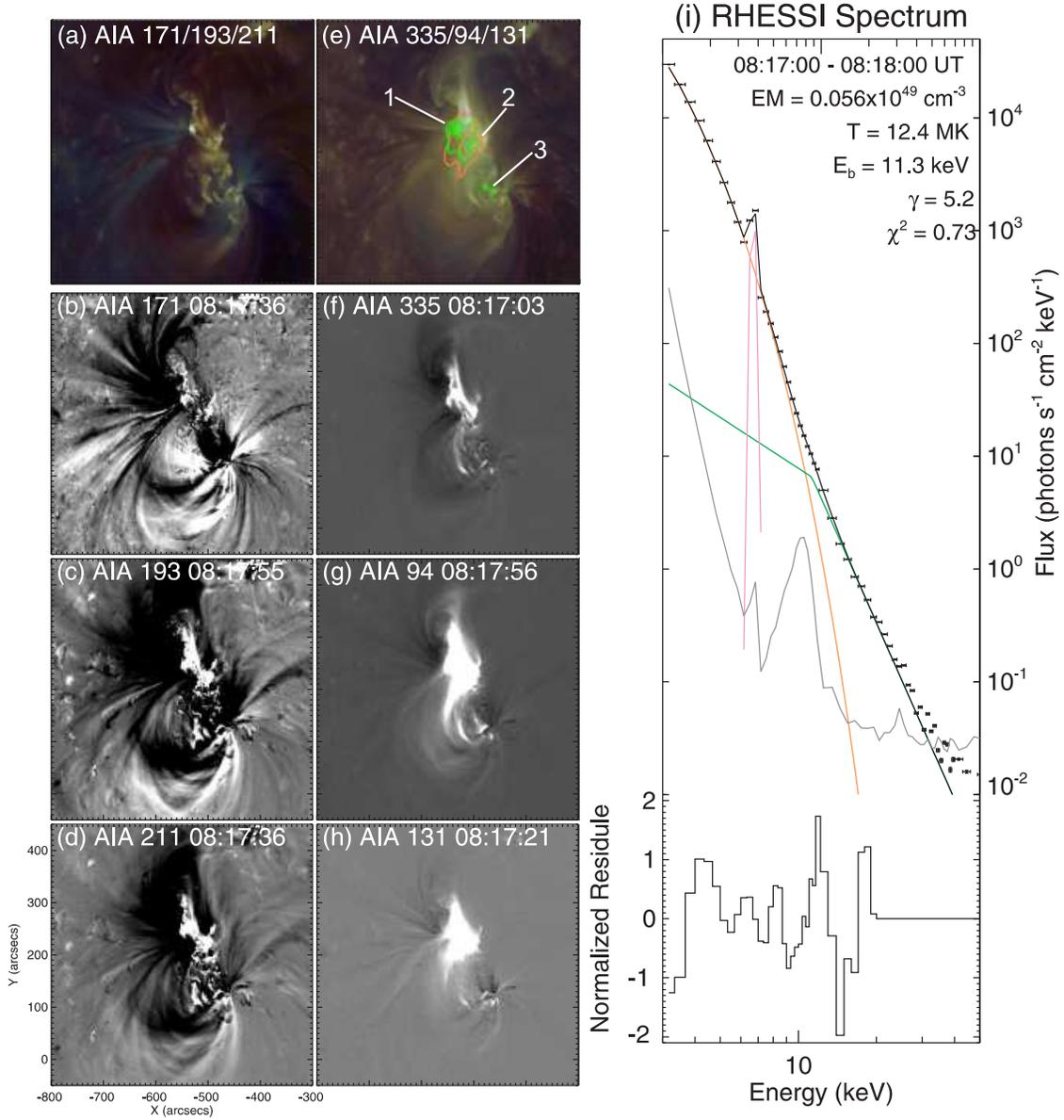} \caption{Flare observed by AIA and \sat{rhessi}. Panels (a) and (e) show
composite images of cold filters (171~\AA\ , 193~\AA\ , and 211~\AA) and hot filters (335~\AA\ ,
94~\AA\ , and 131~\AA), respectively. Panels (b)--(d) and (f)--(h) show the base-difference image
of each individual filter (subtracted by an image at about 06:00 UT). Panel (e) is overlaid by the
same HXR sources as in \fig{cme}(g), with three 12--25 keV sources being marked. Panel (i) shows a
\sat{rhessi} spectrum, with fit residuals plotted at the bottom (see the text for spectral fitting
details). \label{flare}}
\end{figure}

\sat{rhessi} provides additional diagnostic information on the flare loops. An HXR spectrum
integrated from 08:17 UT to 08:18 UT is shown in \fig{flare}(i) as an example. It can be well
fitted with an isothermal component of 12 MK (orange color) plus a power-law component with
$\gamma\simeq5$ above 11 keV up to 25 keV (green). A Gaussian line (purple) is adopted to fit the
iron-line complex peaking at $\sim\,$6.7 keV. The HXR sources reconstructed with the CLEAN
algorithm are shown as contours in \fig{flare}(e). One can see that the thermal source (3--9 keV;
orange contours) is co-spatial with the brightening in hot AIA filters, 94~\AA\ (\fig{flare}(g))
and 131~\AA\ (\fig{flare}(h)). The nonthermal component (12--25 keV; green contours) exhibits two
strong sources: one is apparently associated with Ribbon P (Source 1) and the other, which is also
registered with positive polarity when projected onto the magnetogram (see also \fig{cme}(g)), is
co-spatial with the thermal source (Source 2). Hence we conclude that Source 1 is a footpoint
source and Source 2 is a looptop source. There is an additional weak source (Source 3) co-spatial
with Ribbon N2 (\fig{cme}(g)), which appears to the be the footpoint conjugate to Source 1. Before
08:05 UT, only Source 1 was visible (not shown). We suspect that its conjugate footpoint at Ribbon
N1 is too weak to be detected with \sat{rhessi}'s limited dynamic range ($\sim\,$1:10).

In addition, significant dimmings can be seen in cold AIA filters (\fig{flare}(b)--(d)), but not in
hot filters (\fig{flare}(f)--(h)). This implies that coronal dimming is not only an effect of
density depletion with the cold envelope filed being opened up, but also a temperature effect with
plasmas in the core region being heated to high temperatures.

\section{Discussion \& Conclusion}

The sigmoid eruption on 2010 August 1 evolved clearly in a sequential manner, which provides us an
opportunity to address several controversial issues in solar physics.

First of all, the observation verifies, for the first time, that the topological reconfiguration
due to tether-cutting reconnection does exist in the solar corona, i.e., two opposite J-shaped
loops reconnect to form a continuous S-shaped loop and a compact low-lying loop (\fig{diff}). A
causal relationship between photospheric flows and coronal tether-cutting reconnections is also
evidenced by the detection of persistent converging flows toward the PIL (\fig{flow}). On the other
hand, the observation that the filament was left intact during the sigmoid eruption (\fig{cme}),
which is at variance with what the tether-cutting model predicts, calls for further investigation
on how filaments are embedded in sigmoids.

Second, the observation that the S-shaped loop continuously transformed into an arch-shaped loop
(\fig{diff}) and then erupted as a loop-like CME (\fig{cme}) shed light on the origin of
interplanetary magnetic clouds, which possess rotating magnetic fields \citep{burlaga81}. It also
makes it difficult for one to argue that the observed S-shaped brightening is due to current
density enhancement below a dynamic flux rope.

Third, the observations that the tether-cutting reconnection was not detected as a flare and that
the newly formed flux rope remained in quasi-equilibrium in the lower corona for about 50 minutes
(\fig{diff} and \fig{lc}) argue strongly for the existence of flux ropes prior to solar eruptions
\citep[see also][]{tripathi09, gk09}. We suggest that there may exist a continuum distribution of
the lifetime of ``quiescent'' flux-ropes in the corona from minutes to hours, or even to days,
depending on the amount of magnetic twist possessed by the flux rope and the strength of magnetic
confinement imposed by the surroundings.

Finally, we infer from the observations that most flare loops were formed by reconnection of the
stretched legs of less-sheared J-shaped loops that enveloped the rising flux rope (\fig{diff},
\fig{lc} and \fig{flare}). The reconnection apparently began in the sheared field above the
filament so that some of the sigmoidal field was trapped under the flare arcade, consistent with
the active region remaining sigmoidal during and after the eruption \citep[e.g.,][]{pevtsov02,
mc08}.


\acknowledgments SDO is a mission for NASA's Living With a Star (LWS) Program. KSO H$\alpha$ data
are provided through the Global H-alpha Network operated by NJIT. The authors thank P.~Schuck for
providing the DAVE code. R.L., C.L., S.W. and H.W were supported by NASA grants NNX08-AJ23G and
NNX08-AQ90G, and by NSF grants ATM-0849453 and ATM-0819662. N.D. was supported by NASA grant
NNX08AQ32G.


\end{document}